# Optically driven thermodynamic transition from free- to locked-epitaxy

*Renhong Liang, Mao Ye, Yiran Ying, Longlong Shu, Renkui Zheng, Haitao Huang, Jianhua Hao[*], Shuk-Yin Tong, Shanming Ke[*]*

R.H. Liang, M. Ye, R.K. Zheng, S.M. Ke
School of Physics and Materials Science, Guangzhou University, Guangzhou 510006, China
Email: ksm@gzhu.edu.cn

Y.R. Ying, H.T. Huang, J.H. Hao
Department of Applied Physics, The Hong Kong Polytechnic University, Hong Kong, China.
Email: jh.hao@polyu.edu.hk

L.L. Shu
School of Materials Science and Engineering, Nanchang University, Nanchang 330031, China

S.-Y. Tong
School of Science and Engineering, The Chinese University of Hong Kong (Shenzhen), Shenzhen 518172, China. School of Materials Science and Engineering, Suzhou University of Science and Technology, Suzhou 215009, China

**Abstract**

Controlling crystallographic orientation in quasi–van der Waals (vdW) epitaxy remains a fundamental challenge, especially for material systems located near the boundary between weakly and strongly coupled growth regimes. In such marginal systems, epitaxial selection is governed by a delicate thermodynamic competition between surface-energy penalties and interfacial interaction gains, giving rise to two archetypal limits: vdW-dominated free-epitaxy and strong interfacial coupling dominated locked-epitaxy. However, dynamically driving transitions between these regimes has remained elusive. Here, we demonstrate that external light irradiation can deterministically induce such a transition. Using the thermodynamically frustrated Fe$_4$N/mica interface as a model system, we show that photo-excited carriers act as a chemical potentiator, significantly enhancing the interfacial chemical affinity. Within a quantitative thermodynamic description, this optical modulation increases the locking criterion ($I_{lock}$)—defined as the ratio of interfacial energy gain to surface-energy cost—beyond its critical threshold. As a result, the system switches from vdW-dominated free-epitaxy with (001) orientation to chemically locked-epitaxy with (111) orientation. Our findings establish light as a non-invasive and switchable control knob to dynamically reconfigure the interfacial energy landscape in quasi-vdW epitaxy, enabling programmable access to distinct epitaxial states beyond intrinsic material limitations.

Renhong Liang and Mao Ye contributed equally to this work.

## 1. Introduction

Van der Waals (vdW) epitaxy, initially developed for layered two-dimensional (2D) materials, exploits weak interlayer interactions to relax the stringent lattice-matching constraints inherent to conventional chemically bonded epitaxy. [1–3] More recently, this concept has been extended to the hetero-integration of 2D materials onto three-dimensional (3D) substrates—and vice versa—giving rise to the paradigm of quasi-vdW epitaxy. [4–11] This emerging approach decouples crystallographic compatibility from functional integration. In particular, quasi-vdW epitaxy enables the combination of high-performance bulk materials—such as metals, oxides, nitrides—with low-dimensional substrates that provide reduced interfacial constraint and mechanical compliance. [12–16] Such heterostructures offer a versatile platform for flexible and/or transferable electronics/optoelectronics that are difficult to realize using conventional rigid substrates. [17-22]

Despite these advances, regulating the epitaxial interface in quasi-vdW systems presents a fundamental duality between two limiting regimes. In the free-epitaxy regime, the overlayer interacts weakly with the substrate through vdW interactions, enabling the film to minimize its own surface energy. This mode naturally enables strain relaxation and mechanical peelability, which are highly desirable for flexible and transferable devices. [23-25] However, the absence of a strong orientational driving force leads to rotational degeneracy and limits crystallographic coherence, as schematically illustrated in **Figure 1**a. In contrast, locked-epitaxy emerges when strong interfacial interactions—primarily electrostatic and chemical in nature—dominate the energy landscape. In this regime, the film is forced into a specific lattice registry, yielding high-quality single-crystalline or single-domain alignment. [13,14] Yet this comes at the cost of increased interfacial strain and a loss of the weakly bonded nature that underpins vdW-based transferability. These two regimes therefore represent competing advantages, and the ability to reversibly switch between them would offer unprecedented control over epitaxial functionality. However, in most material systems, the interfacial interaction strength is intrinsically fixed, rendering such tunability largely inaccessible.

Fundamentally, the transition between free- and locked-epitaxy in quasi-vdW systems reflects a thermodynamic competition between the energetic cost of forming a specific crystal facet and the energetic gain arising from interfacial interactions. [26] In realistic growth environments, epitaxial behavior may be influenced by both intrinsic interfacial interactions—such as vdW forces, electrostatic attraction, and site-specific chemical bonding—as well as extrinsic factors including step edges, defects, and kinetic barriers. [27-30] While these extrinsic effects can affect nucleation pathways and domain evolution, the intrinsic interfacial interactions define the thermodynamic landscape that ultimately constrains the accessible epitaxial states.

In this work, we achieved the transition from free-epitaxy to locked-epitaxy via a highly straightforward light irradiation process. Furthermore, within a thermodynamic framework, we elucidated how light irradiation modifies the intrinsic interfacial interactions. To visualize the interfacial competition, we construct a conceptual phase space described by two key descriptors: interfacial polarity coupling, reflecting electrostatic interactions across the interface, and chemical affinity, capturing the propensity for directional chemical bonding between film and substrate atoms (Figure 1b). Based on our thermodynamic model, [26] the locked-epitaxy of the ZnO/mica heterojunction is governed by its interfacial polarity coupling; in contrast, for the $Fe_4N$/mica system, the electrostatic effect is negligible and the chemical interaction is insufficient to sustain "locking", thereby preserving free-epitaxy. Here, we demonstrate that external light irradiation can dynamically modify the interfacial chemical interaction landscape of $Fe_4N$/mica heterostructure. By optically enhancing the interfacial chemical affinity, the system is driven across the phase boundary from free- to locked-epitaxy, realizing a deterministic and reversible control of epitaxial orientation. This strategy transforms quasi-vdW epitaxy from a passive thermodynamic outcome into an actively tunable, optically programmable process.

## 2. Results and Discussion

To demonstrate the concept of light-controlled epitaxial regulation, we employed a light-assisted sputtering configuration, schematically illustrated in Figure 1 (see also Figure S1, where the incident light directly illuminates the mica substrate through an optical window). In the absence of light irradiation, both the deposited species and the mica surface interact weakly (Figure 1c). Under these conditions, epitaxial growth is governed primarily by vdW interactions, enabled by the chemically stable surface of mica and the weak, non-directional adatom-substrate coupling. As a result, the interface provides little orientational guidance for the overlayer, and the system is unable to overcome the intrinsic energetic penalty associated with forming high-energy facets. Consequently, the system relaxes into a thermodynamically frustrated but globally stable state, adopting the lowest-energy $Fe_4N$ (001) orientation. This leads to a symmetry-mismatched interface (Figure 1c), where the fourfold-symmetric $Fe_4N$ (001) film interacts weakly with the sixfold-symmetric mica (001) substrate through isotropic vdW forces, characteristic of free-epitaxy.

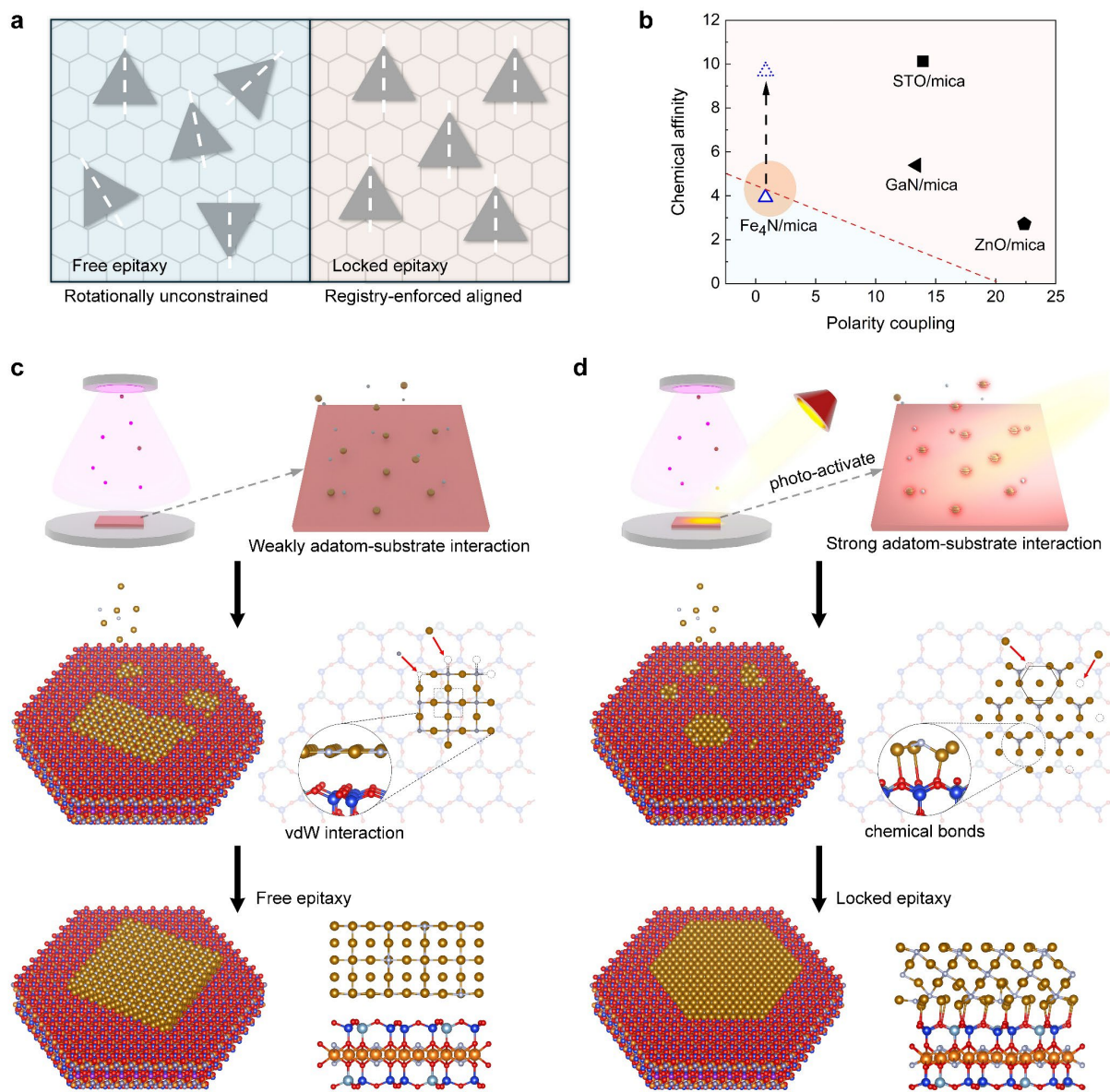

**Figure 1. Schematic illustration of the optically-driven transition from free- to locked-epitaxy in the Fe₄N/mica heterostructure.** (a) Schematic of free- versus locked-epitaxy, illustrating rotationally unconstrained growth versus registry-enforced alignment at quasi-vdW interfaces. In 3D/2D cases, there is also the situation of out-of-plane locking. (b) Conceptual phase diagram mapping representative 3D/mica systems as a function of interfacial polarity coupling and chemical affinity. The red dashed line denotes the thermodynamic boundary separating free- and locked-epitaxy. (c) In the absence of light irradiation, epitaxial growth is governed by surface energy minimization. The Fe₄N film adopts its lowest-energy (001) facet with 4-fold symmetry, forming a symmetry-mismatched, rotationally disordered interface with the 6-fold mica substrate dominated by weak vdW interactions. (d) Under light irradiation, photo-excited carriers enhance the interfacial chemical affinity, driving the system across the locking boundary. As a result, the Fe₄N film switches to the (111) orientation to maximize lattice registry with mica, forming a symmetry-matched, chemically locked interface.

Interestingly, this thermodynamic balance can be dynamically altered by external light irradiation. The mica substrate is transparent in the visible range (Figure S2), allowing incident photons to selectively interact with the adsorbed species at the growth front. Such optical excitation is known to generate excess photocarriers, [31-34] which can transiently populate interfacial electronic states. We propose that these photo-excited carriers effectively act as a chemical potentiator, selectively amplifying the bonding tendency between the adsorbate species and the mica surface. Rather than modifying the intrinsic surface energetics of $Fe_4N$, light irradiation enhances the interfacial interaction strength, thereby increasing the energetic gain associated with forming a symmetry-matched interface. As a consequence, the balance between surface energy cost and interfacial coupling is fundamentally reshaped: the interfacial driving force becomes sufficiently strong to compete with the otherwise prohibitive surface energy penalty of the $Fe_4N$ (111) facet. The system is thus driven away from the free epitaxial state and forced to adopt the (111) orientation (Figure 1d), maximizing atomic registry with the hexagonal lattice of mica and realizing a locked-epitaxy state.

The crystallinity and epitaxial orientation of the $Fe_4N$ films were systematically characterized by X-ray diffraction (XRD) measurements (**Figure 2** and Figure S3). For films deposited in the absence of light, the $\theta$-$2\theta$ scan reveals that $Fe_4N$ grows preferentially along the (001) orientation, consistent with previous reports on $Fe_4N$/mica heterostructures from different research group. [35-37] In contrast, when light irradiation is introduced during sputtering, a clear orientation transition occurs: the dominant diffraction peaks correspond to $Fe_4N$ (111), indicating a light-induced switch in the out-of-plane growth direction. This structural transition is further corroborated by cross-sectional TEM analysis (Figure S4). Rocking curve measurements around the $Fe_4N$ (001) and $Fe_4N$ (111) reflections yield narrow full widths at half maximum (FWHM) of 0.8° and 0.75°, respectively, demonstrating that both films possess high crystalline quality despite their distinct epitaxial modes. Therefore, light irradiation does not degrade crystalline quality but selectively alters the epitaxial selection.

To elucidate the in-plane epitaxial relationship, we first examine the light-induced $Fe_4N$ (111)/mica heterostructure. A $\varphi$-scan around the $Fe_4N$ (131) reflection ($2\theta$=84.7°, $\chi$=0°), together with a pole figure measured around the $Fe_4N$ (111) family ($2\theta$=41.2°, $\chi$=0-75°), reveals a pronounced six-fold symmetry (Figure 2g, h). This directly reflects a well-defined in-plane registry between the $Fe_4N$ (111) film and the hexagonal mica substrate, corresponding to the locked-epitaxy regime illustrated in Figure 1. Reciprocal space mapping (RSM) measurements further clarify the strain state of the $Fe_4N$ (111) film (Figure S5). The symmetric RSM shows that the $Fe_4N$ (111) diffraction peak aligns vertically with the mica (005) peak, indicating full relaxation along the out-of-plane direction. From the asymmetric RSM around the $Fe_4N$ (132) reflection, the in-plane lattice constant is extracted to be $a$=3.733 Å, corresponding to a compressive strain of ~1.45%. This finite in-plane strain, together with the unique in-plane orientation, is characteristic of a locked-epitaxy mode, where strong interfacial coupling enforces crystallographic registry.

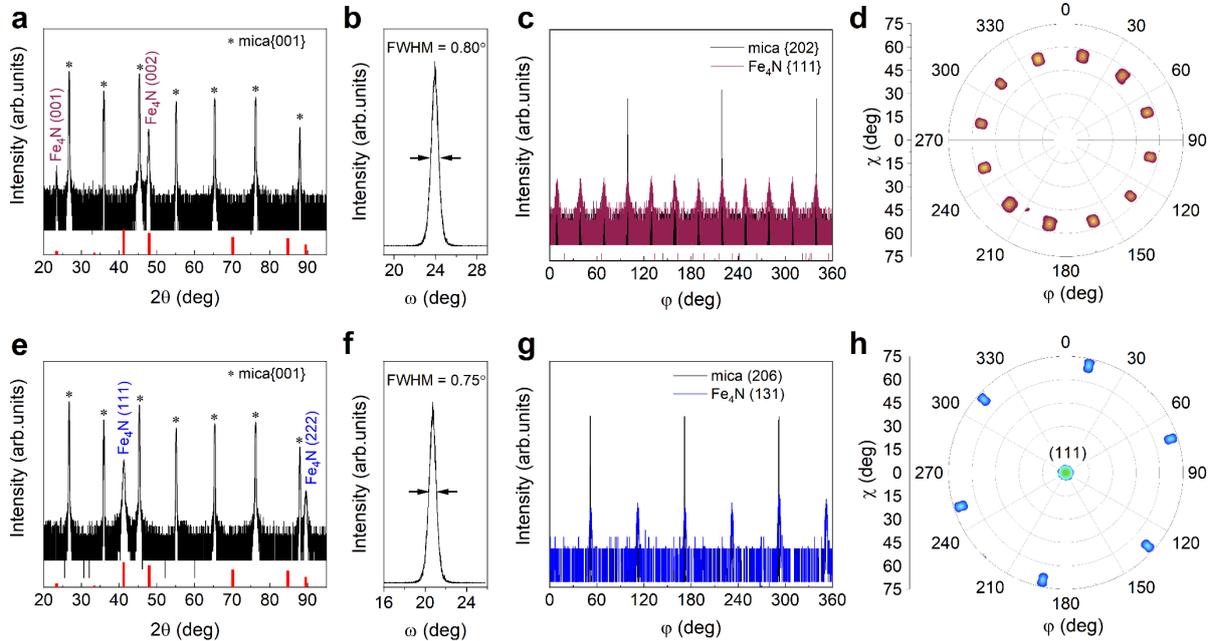

**Figure 2. Structural characterization of Fe$_4$N/mica heterostructures by X-ray diffraction.** XRD *θ-2θ* scans of Fe$_4$N films grown on mica under identical sputtering conditions (a) without light irradiation and (e) with light irradiation. In the absence of light, the Fe$_4$N film adopts a (001)-oriented growth mode, whereas light irradiation induces a clear transition to a (111)-oriented epitaxial growth. Rocking curves of the (b) Fe$_4$N (001) and (f) Fe$_4$N (111) reflections exhibit narrow FWHM, indicating high crystalline quality in both growth modes. In-plane epitaxial relationships are examined by *φ*-scan measurements for (c) Fe$_4$N (001) and (g) Fe$_4$N (111) films. Corresponding pole figures measured around the Fe$_4$N {111} reflections for (d) Fe$_4$N (001) and (h) Fe$_4$N (111) films further reveal rotationally degenerate domains in the free epitaxy regime and a well-defined six-fold symmetry in the locked epitaxy regime.

In contrast, the Fe$_4$N (001) film grown without light irradiation exhibits hallmarks of free-epitaxy. The lattice constant extracted from XRD is $a=3.795$ Å, in excellent agreement with the bulk value, indicating negligible strain. Moreover, *φ*-scan measurements show a 12-fold symmetry with peaks separated by 30° (Figure 2c). Consistently, 12 poles spaced by 30° are observed at a tilt angle of $\chi=54.7°$ in the pole figure (Figure 2d). This diffraction pattern arises from the rotational symmetry mismatch between the four-fold cubic Fe$_4$N (001) film and the six-fold mica substrate, resulting in three equivalent rotational domains and a rotationally degenerate interface (Figure S6).[38] Such behavior is a defining feature of vdW-mediated free-epitaxy.

The above results clearly demonstrate that light irradiation can induce a transition in the Fe$_4$N/mica system from free-epitaxy to locked-epitaxy. To establish the intrinsic thermodynamic barrier for this transition, we first quantify the surface energy cost associated with different Fe$_4$N facets. As shown in **Figure 3**a, the low-index Fe$_4$N (001) surfaces exhibit substantially lower formation energies than the Fe$_4$N (111) terminations. The resulting surface-energy difference ($\Delta\gamma_{sur}$~0.03 eV Å$^{-2}$) represents a rigid thermodynamic penalty for adopting the (111) orientation. This penalty is an intrinsic property of the Fe$_4$N crystal and is independent of external stimuli, thereby strongly favoring (001)-oriented growth under equilibrium conditions.

We next examine how external light irradiation acts on this intrinsically unfavorable energetic landscape. Importantly, light does not modify the surface energies of Fe$_4$N facets, nor does it alter the lattice geometry of the film. Instead, its effect is confined to the interfacial interaction. we parameterize the Fe$_4$N/mica interface using two physically motivated descriptors: polarity coupling and chemical affinity. The polarity coupling captures the electrostatic interaction between the film and substrate and is approximated by the product of their intrinsic surface potential steps ($\Delta V_{film} \times \Delta V_{sub}$), extracted from planar-averaged electrostatic potential profiles (Figure S7). The chemical affinity reflects the tendency for site-specific bonding at the interface and is quantified by the adsorption energy of adatoms on the substrate surface (Figure S8). Detailed definitions and computational procedures are provided in the Supporting Information. Within this parameterized polarity–affinity phase space (Figure 3b), the Fe$_4$N/mica system resides in a marginal free-epitaxy regime under dark conditions, reflecting its limited interfacial reactivity. Upon illumination, photo-excited carriers are generated and populate interfacial bonding states, as evidenced by the markedly enhanced adsorption strength of Fe adatoms on mica. This process selectively amplifies the chemical affinity component, shifting the system vertically toward the locked regime without altering its intrinsic surface-energy cost.

To quantitatively determine whether the photo-enhanced interfacial interaction can overcome the intrinsic surface-energy barrier, we decompose the total energetics of the Fe$_4$N/mica interface. As shown in Figure 3c and Table S1, the total energy balance is resolved into an energy cost—comprising the surface energy penalty ($\Delta\gamma_{sur}$), vdW interaction ($\Delta\gamma_{vdW}$), and lattice strain ($\Delta E_{strain}$)—and an interfacial gain arising from electrostatic ($\Delta\gamma_{es}$) and chemical bonding ($\Delta\gamma_{chem}$) contributions. In the dark state, the interfacial interaction is dominated by vdW coupling, while the chemical contribution remains modest. Consequently, the total interfacial gain is insufficient to compensate the large surface-energy cost associated with the (111) facet, as summarized in the grouped comparison in Figure 3d. Under light irradiation, the energetic landscape is fundamentally reconfigured. While the energy cost remains essentially unchanged, the chemical bonding term $\Delta\gamma_{chem}$ is dramatically enhanced, leading to a substantial increase in the total interfacial gain. In the compressed representation, this photo-enhanced gain decisively exceeds the surface-energy penalty. This balance inversion is captured by the locking criterion, $I_{lock}$=(interfacial gain)/(energy cost). As plotted in Figure 3e, $I_{lock}$ increases from below unity in the dark state to approximately 1.4 under illumination, crossing the critical threshold for epitaxial locking. [26] This transition marks a shift from a thermodynamically frustrated free-epitaxy regime to a strictly locked state, in which the system is compelled to adopt the (111) orientation to maximize interfacial registry.

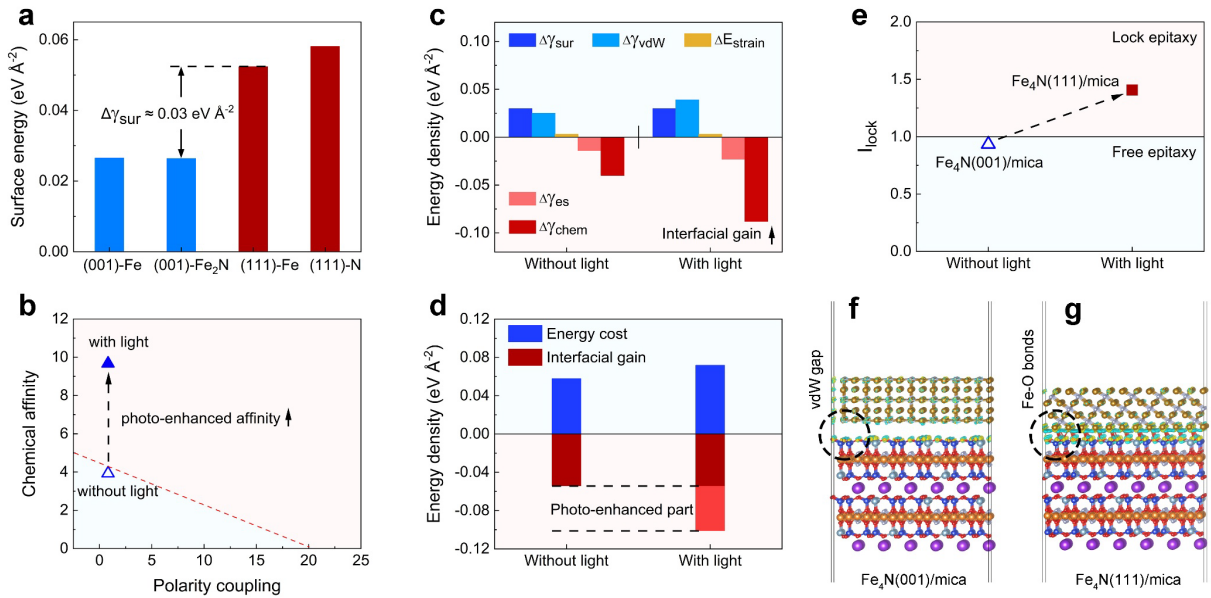

**Figure 3. Thermodynamic decomposition of the photo-induced free-to-locked-epitaxy transition in Fe₄N/mica.** (a) Surface formation energies of Fe₄N facets with different terminations, highlighting the substantial intrinsic surface-energy penalty associated with forming the (111) orientation relative to the low-energy (001) facet. This surface-energy difference defines a rigid thermodynamic barrier that disfavors (111) growth under equilibrium conditions. (b) Position of the Fe₄N/mica system in a polarity–affinity phase space under dark and illuminated conditions. Light irradiation selectively enhances the interfacial chemical affinity, shifting the system vertically from the marginal free-epitaxy region into the locked regime without altering its intrinsic surface energetics. (c) Detailed decomposition of the interfacial energetics into energy costs (surface energy penalty $\Delta\gamma_{sur}$, vdW interaction, and lattice strain) and energy gains (electrostatic $\Delta\gamma_{es}$ and chemical bonding $\Delta\gamma_{chem}$ contributions) for Fe₄N/mica interfaces with and without light irradiation. The blue and red backgrounds represent the energy cost and interfacial gain, respectively. (d) Grouped comparison of the total energy cost and interfacial gain, illustrating the inversion of the energetic balance upon illumination: while the gain is insufficient to overcome the surface-energy penalty in the dark state, photo-enhanced chemical interactions dominate under light. (e) Evolution of the locking criterion, $I_{lock}$ = (interfacial gain)/(energy cost), showing a sharp transition across the critical threshold ($I_{lock}$ = 1) from free epitaxy to locked epitaxy under illumination. (f, g) Charge

density difference visualizations for Fe$_4$N (001)/mica and Fe$_4$N (111)/mica interfaces, respectively. The (001) interface exhibits a vdW-like gap with minimal charge redistribution, whereas the (111) interface under illumination displays pronounced charge accumulation and directional Fe–O bonding, providing the microscopic origin of epitaxial locking.

Charge density difference analysis provides microscopic confirmation of this transition (Figure 3f, g). The Fe$_4$N (001)/mica interface exhibits a clear vdW gap with minimal charge redistribution, characteristic of weak coupling. In contrast, the illuminated Fe$_4$N (111)/mica interface displays pronounced charge accumulation and orbital hybridization between Fe and surface O atoms, forming strong directional Fe-O bonds (~1.98 Å), comparable to those in α-Fe$_2$O$_3$.[39] These bonds serve as the atomic-scale anchor that rigidly locks the epitaxial lattice to the substrate, enforcing the six-fold symmetry observed experimentally.

Guided by the above mechanism, the magnitude of the interfacial gain should be tunable controlled by the characteristics of the light stimuli. We systematically investigated the impact of irradiation timing, wavelength, and intensity to map the boundaries of this transition. First, the timing of irradiation reveals that the crystallographic registry is determined at the very onset of growth. As shown in Figure S9, light irradiation is critical only during the initial deposition stage (nucleation and early coalescence). Once the locked (111) template is established, subsequent growth maintains this orientation even if the light is switched off. This confirms that the thermodynamic competition between surface energy and interfacial coupling occurs primarily during the nucleation phase,[31] where the surface-to-volume ratio is highest. Second, the wavelength dependence provides decisive evidence for the photo-carrier mechanism over a photothermal one. As shown in Figure S10, both white light and 365 nm UV light successfully drive the transition from free-(001) to locked-(111). In sharp contrast, 850 nm irradiation fails to induce any locking, resulting in the (001) orientation identical to the dark state. This cutoff correlates perfectly with the optical absorption properties: the photon energy of 850 nm light (~1.46 eV) is below the optical bandgap of the Fe$_4$N adsorbate (Figure S11), leading to negligible carrier generation. This confirms that the enhancement of the chemical affinity relies strictly on the population of photo-excited carriers, rather than local heating.

Moreover, the light intensity dependence visualizes the continuous trajectory of the transition. As shown in Figure S12, we observe a clear intensity-dependent phase boundary. When the light intensity is weak, the system remains in the free regime and shows (001)-dominant due to insufficient carrier concentration. At intermediate intensities, a coexistence of (001) and (111) domains emerge. This signifies that the system is hovering exactly at the critical thermodynamic threshold ($I_{lock}$~1), where the interfacial gain roughly balances the surface energy cost. When beyond a critical intensity, the orientation undergoes a complete transition to the lock (111) state.

The thermodynamic transition from free-to locked-epitaxy dictates not only the crystallographic orientation but also the macroscopic growth mode. According to the classical Bauer's criterion, the growth mode is governed by the surface energy balance. In the dark state, the system selects the $Fe_4N$ (001) facet due to its low surface energy cost. This thermodynamic preference facilitates wetting, promoting a Frank-van der Merwe (layer-by-layer) growth mode. Indeed, AFM topography reveals that the $Fe_4N$ (001) films maintain a remarkably smooth and continuous morphology even with increasing sputtering time (**Figure 4**a). Conversely, under illumination, the system is forced into the $Fe_4N$ (111) orientation. Despite the strong interfacial bonding, the high surface energy of the (111) facet drives the film to minimize surface area by clustering, triggering a transition to the Volmer-Weber (island) growth mode. As shown in Figure 4b, the $Fe_4N$ (111) films exhibit discrete grains that coarsen significantly over time (average grain size increases from 8 nm at 30 s to 18 nm at 3 min), standing in sharp contrast to the 2D morphology of the dark-state samples.

To further verify the crystal quality and surface flatness, we performed ex-situ reflection high-energy electron diffraction (RHEED). Figure 4c schematically illustrates the RHEED geometry, where the electron beam strikes the surface at a grazing incidence to probe the near-surface structure. For the dark-state samples (Figure 4d), the RHEED patterns of $Fe_4N$ (001) (Figure $4d_3$-$d_4$) exhibit sharp, vertical streaks, closely resembling those of the atomically flat mica substrate (Figure $4d_1$-$d_2$). This persistence of streaky patterns is a hallmark of 2D layer-by-layer growth, confirming that the (001) film grows flat without forming 3D protrusions. In contrast, the light-assisted $Fe_4N$ (111) samples (Figure 4e) display distinct spotty transmission patterns (Figure $4e_3$-$e_6$). These spots arise from the electron beam penetrating through the 3D islands (surface asperities), providing definitive reciprocal-space evidence for the island growth mode. This combination of real-space (AFM) and reciprocal-space (RHEED) analysis confirms that the light-driven thermodynamic transition is accompanied by a fundamental switch in growth kinetics: from surface-energy-minimized 2D flow to interface-energy-maximized 3D locking.

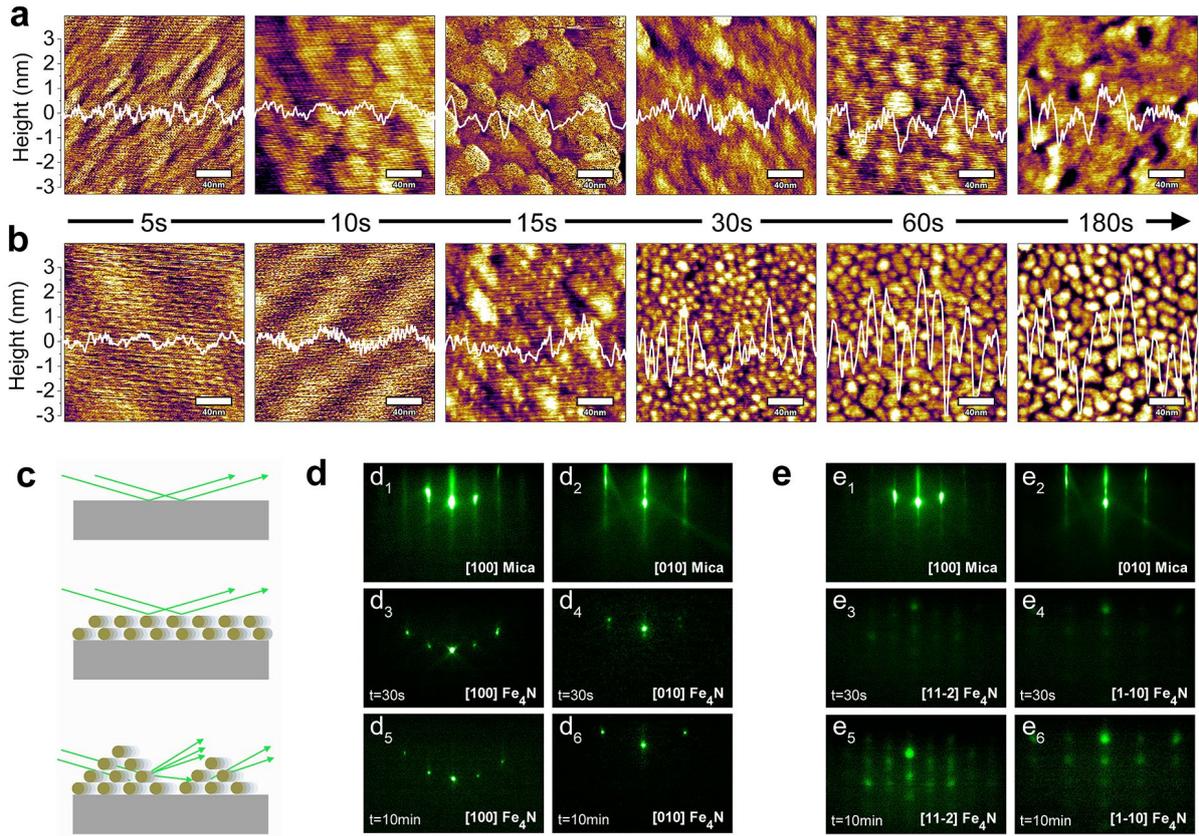

**Figure 4. Distinct growth modes driven by thermodynamic selection.** (a, b) AFM topographic evolution of Fe$_4$N films with increasing deposition time (from left to right). (a) Fe$_4$N (001) exhibits a smooth surface consistent with layer-by-layer (Frank-van der Merwe) growth. (b) Fe$_4$N (111) displays granular coarsening, indicative of island (Volmer-Weber) growth driven by high surface energy. The white curves in each figure represent the height fluctuation of the morphology. The scale bar is 40 nm. (c) Schematic illustration of the RHEED geometry. (d, e) RHEED patterns corresponding to the films in (a) and (b). (d) Fe$_4$N (001)/mica shows streaky patterns characteristic of a flat 2D surface. (e) Fe$_4$N (111)/mica shows transmission spots confirming the 3D island morphology. Patterns are shown for bare mica (d$_1$, d$_2$ and e$_1$, e$_2$), 30 s deposition (d$_3$, d$_4$, e$_3$, e$_4$), and 10 min deposition (d$_5$, d$_6$, e$_5$, e$_6$).

## 3. Conclusion

In summary, our work establishes that the epitaxial ground state in marginal quasi-vdW systems is not an immutable material property, but a dynamically tunable thermodynamic parameter. By utilizing light excitation to enhance the chemical affinity, we successfully drove the thermodynamically frustrated Fe$_4$N/mica system across the critical locking boundary, signifies that interfacial gain outweighs surface-energy cost. This transition enables a deterministic switch from vdW-dominated free-epitaxy to chemically locked-epitaxy. Compared with conventional thermal or compositional tuning strategies, the optical approach offers distinct advantages. As a non-contact stimulus, light can be applied uniformly or patterned spatially. By controlling the illumination profile (e.g., using optical masks or baffles), locked and free epitaxial regions can be selectively programmed on the same substrate, enabling spatially resolved control over crystalline orientation and functional properties

Crucially, this study represents the complementary counterpart to our recent findings on the SrTiO$_3$/mica system, [40] where electrostatic polarity was deliberately screened by defect engineering to reduce the locking criterion and store the free vdW growth mode. Together, these two studies delineate a unified roadmap for interface engineering in quasi-vdW epitaxy: by independently modulating chemical reactivity (via light) or electrostatic polarity (via defects), one can precisely navigate the quasi-vdW phase diagram in either direction. One pathway increases $I_{lock}$ to enforce rigid epitaxial locking for high-quality, single-crystalline integration, while the other decreases $I_{lock}$ to recover vdW-like behavior, enabling peelability and transferability without sacrificial layers. Ultimately, this work transforms quasi-vdW epitaxy from a passive consequence of interfacial forces into an actively controllable platform, opening new avenues for the rational design of heterogeneous electronic, magnetic, and photonic devices.

## 4. Experimental Method

**Thin film epitaxial growth:** Fe$_4$N films were deposited on mica substrate by DC magnetron sputtering system (Kurt J. Lesker). The Fe target has a diameter of 50 mm, purity of 99.95%. The target-to-substrate distance was set at 100 mm. The base pressure was lower than 1×10$^{-7}$ Torr. The Fe target was pre-sputtered for 10 min prior to Fe$_4$N film deposition. The substrate temperature was kept at 450 °C, and to obtain accurate stoichiometric Fe$_4$N thin films, an Ar (99.999%) and N$_2$ (99.999%) gas mixture with a mixing ratio of Ar:N$_2$=5:1 was introduced into the chamber. The total pressure was 8.5 mTorr, and the sputtering power on the target was 38 W. A commercial lamp with changeable power (up to 100 W) and wavelength is placed outside the chamber window.

**Characterizations:** The crystal structure and epitaxial relationship of the films were measured by High resolution X-ray diffraction (XRD, Rigaku smartlab) with Cu Kα radiation (λ = 1.5406 nm) and Reflection High Energy Electron Diffraction (RHEED). High-resolution transmission electron microscopy (HRTEM) images were recorded at 200 kV using a Tecnai G2 F20 S-Twin instrument. The specimens for cross-sectional TEM observation of the mica-based heterostructures were prepared by using a focused ion beam (FIB) technique. The surface morphology and potential were determined using atomic force microscopy (AFM, Cypher ES, Asylum Research) in tapping mode with a silicon tip cantilever (AC240TS, Oxford) and Kelvin

Probe Force Microscopy (KPFM) mode with a conductive tip cantilever (ASYELEC.01-R2), respectively.

**First-principle calculations:** All density functional theory (DFT) calculations were performed using the Quickstep module of the CP2K package, employing a mixed Gaussian and plane-wave (GPW) approach. [41] The inner core electrons were represented with norm-conserving Goedecker-Teter-Hutter (GTH) pseudopotentials. [42] For the structural optimizations, the Kohn-Sham orbitals were expanded in a Double-Zeta Valence plus Polarization (DZVP-MOLOPT-GTH) Gaussian-type basis set [43] with a plane wave cutoff of 400 Ry for the finest level of the multi-grid. For energy calculations, a more complete Triple-Zeta Valence plus Double Polarization (TZV2P-MOLOPT-GTH) basis set was employed, combined with an increased plane wave cutoff of 600 Ry. Geometry optimizations and binding energy calculations utilized the generalized gradient approximation (GGA) in the form of the Perdew-Burke-Ernzerhof (PBE) exchange-correlation functional. [44] To account for long-range van der Waals dispersion interactions between the films and mica substrate, the DFT-D3(BJ) correction scheme [45] was incorporated into all energy evaluations.

The initial bulk structures of $Fe_4N$ and mica were fully optimized at the DFT level. The optimized lattice constant of bulk $Fe_4N$ was found to be $a$ = 3.774 Å, while those for monoclinic mica were $a$ = 5.347 Å and $b$ = 9.277 Å. Based on these results, we constructed slab models with different surface terminations: $Fe_2N$- and Fe-terminated $Fe_4N$ (001), as well as Fe- and N-terminated $Fe_4N$ (111), and the $(AlSi)_2O_3$-terminated mica(001) surfaces. To minimize the effects of periodic boundary conditions on atomic interactions, we selected supercells consisting of a five-layer of 7×7 $Fe_4N$ (001) sitting on 2 layer of 5×3 mica (26.7 Å×27.2 Å×66.1 Å) for $Fe_4N$ (001)/mica system. Additionally, a supercell of seven-layer of $Fe_4N$ (111) with $a$=$b$=32.02 Å, $\gamma$=120°, and mica (001) with $a$=32.0 Å, $b$=32.0 Å, $\gamma$=120° (32.0 Å×32.0 Å×63.5 Å) was considered for the $Fe_4N$ (111)/mica system. For the calculations, we applied the orbital transformation (OT) method at the Γ-point for both of the $Fe_4N$ (001)/mica and $Fe_4N$ (111)/mica systems.

To investigate the experimentally observed light-induced transition of epitaxial $Fe_4N$ growth from (001) to (111) orientation on mica, we performed additional structure optimizations under the lowest singlet excited state using time-dependent DFT (TDDFT) within the Tamm-Dancoff approximation (TDA) as implemented in CP2K. [46] To evaluate the computational efficiency and enable comparison with ground-state optimizations, these excited-state optimizations were performed under identical computational settings to those used in the ground-state DFT calculations, encompassing the same basis sets, cutoffs, convergence thresholds, and dispersion corrections. This approach yields approximate equilibrium geometries that facilitate the assessment of photoinduced structural changes.

To calculated the surface energy of Fe₄N (001) and Fe₄N (111), we initially calculated the cleavage energy of Fe₄N, $E_{cleavage}$, and distributed it to different terminations. For example, the unrelaxed internal cleavage energy of (001) surface was calculated as [47]

$$E^{(001)}_{cleavage} = \frac{1}{4}\left[E^{(001)-Fe}_{slab} + E^{(001)-Fe_2N}_{slab} - x\Delta E_{bulk}\right] \quad (1)$$

where $E^{(001)-Fe}_{slab}$ and $E^{(001)-Fe_2N}_{slab}$ are the energies of the unrelaxed Fe-terminated and Fe₂N-terminated Fe₄N (001) surfaces, respectively, $\Delta E_{bulk}$ represents the formation energy of a single bulk Fe₄N unit cell, and $x$ is the number of total unit cells in both slabs. The factor of $\frac{1}{4}$ is included as four surfaces were created upon cleavage.

Based on $E^{(001)}_{cleavage}$ and $\Delta E_{bulk}$, the surfaces energy of the Fe-terminated Fe₄N (001) surface can be calculated as

$$E^{(001)-Fe}_{surface} = E^{(001)}_{cleavage} + \frac{1}{2}\Delta E^{(001)-Fe}_{relaxation} \quad (2)$$

where $\Delta E^{(001)-Fe}_{relaxation}$ is the energy released upon the relaxation of the slab.

The binding energy was calculated according to $E_{bind} = E_{film/substrate} - [E_{substrate} + E_{film}]$, where $E_{film/substrate}$ is the total energy of the heterostructure, $E_{substrate}$ and $E_{film}$ are the total energy of the substrate slab and film slab, respectively. Basis-set superposition errors (BSSE) were estimated via the counterpoise correction to improve accuracy.

In our model, the total interface energy ($\gamma_{int}$) is decomposed into chemical ($\gamma_{chem}$), electrostatic ($\gamma_{es}$), and van der Waals ($\gamma_{vdW}$) contributions. Here, $\gamma_{vdW}$ specifically refers to the non-local dispersion forces (as calculated by the DFT-D3 method), while $\gamma_{es}$ encapsulates all other classical electrostatic interactions—including orientation and induction forces—that are captured by the standard DFT functional. To quantify the $\gamma_{es}$ arising from charge rearrangement at the heterostructure interface, we analyzed the planar-averaged charge density difference and its corresponding electrostatic potential. Using the Multiwfn program, [48,49] we calculated and analyzed the planar-averaged charge density difference, $\Delta\rho(z)$, by subtracting the charge densities of the isolated components from the total charge density of the fully relaxed heterostructure:

$$\Delta\rho(z) = \rho_{total}(z) - \rho_{film}(z) - \rho_{substrate}(z) \quad (3)$$

The planar-averaged electrostatic potential, $V(z)$, was obtained by similar method from $\Delta\rho(z)$. The electrostatic energy per unit area ($\gamma_{es}$) stored in this interfacial charge layer is then given by the classical expression for the energy of a charge distribution in its own field [50]

$$\gamma_{es} = \frac{1}{2} \int \Delta\rho(z) \cdot V(z) dz \tag{4}$$

**Conflicts of Interest**

The authors declare no conflict of interest

**Data Availability Statement**

All data needed to evaluate the conclusions in the paper are present in the paper and/or the Supplementary Materials.

**Supporting Information**

Supporting Information is available from the Wiley Online Library or from the author.

**The table of contents**

This study demonstrates a straightforward light-regulated quasi-van der Waals epitaxial interface approach, enabling the successful transition from free- to locked-epitaxy. The adsorbed species on the substrate surface absorb photons to generate photogenerated carriers, which modulate interfacial chemical affinity and thereby enable a transition from van der Waals-dominated free-epitaxy growth to chemical bond-guided locked-epitaxial growth.

**ToC figure**

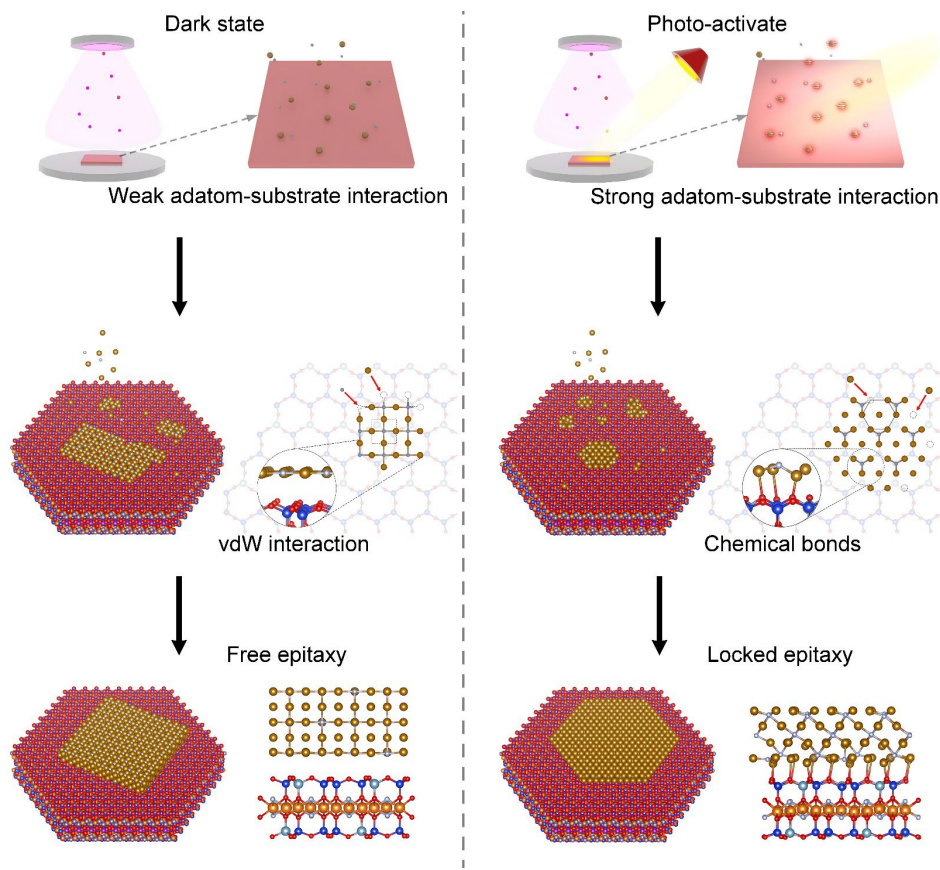